\documentclass[aps,pre,showpacs,twocolumn,groupedaddress]{revtex4}

\usepackage{graphicx,bm,amssymb}

\begin{document}
\title{Free planar isotropic-nematic interfaces in binary hard-rod fluids}
\author{Kostya Shundyak and Ren\'e van Roij}
\affiliation {Institute for Theoretical Physics, Utrecht University,\\
Leuvenlaan 4, 3584 CE Utrecht, The Netherlands}
\date{August 20, 2003}

\begin{abstract}
Within the Onsager theory we study free planar isotropic-nematic
interfaces in binary mixtures of hard rods. For sufficiently
different particle shapes the bulk phase diagrams of these
mixtures exhibit a triple point, where an isotropic (I) phase
coexists with two nematic phases ($N_1$ and $N_2$) of different
composition. For all explored mixtures we find that upon approach
of the triple point the $IN_2$ interface shows complete wetting by
an intervening $N_1$ film. We compute the surface tensions of
isotropic-nematic interfaces, and find a remarkable increase with
fractionation, similar to the effect in polydisperse hard-rod
fluids.
\end{abstract}

\pacs{61.30.Hn, 05.70.Np, 68.03.Cd, 68.08.Bc}

\maketitle

\section{Introduction}
\label{inroduct} Colloidal suspensions of rodlike particles are
well-known to exhibit a wealthy phase behavior as a function of
density \cite{VL,AEFM,SF}. Early experiments on vanadiumpentoxide
solutions \cite{ZZAC25} and suspensions of Tobacco Mosaic Virus
(TMV) particles \cite{BPBFN36} showed a first-order transition
from a disordered (isotropic) fluid phase to an ordered
(liquid-crystalline nematic) fluid phase upon increasing the
concentration of rods sufficiently. The nematic is homogeneous (it
is a fluid), but the rods are on average oriented in a specific
direction $\hat{n}$, the so-called nematic director. The bulk
nematic ordering is of uniaxial symmetry, i.e., there is azimuthal
symmetry about $\hat{n}$.  The isotropic-nematic ($IN$) transition
was first explained by Onsager, who modelled the colloidal rods as
hard needles. He first showed that the average pairwise excluded
volume is reduced in the nematic phase compared to the isotropic
phase, and then argued that the resulting gain of free volume (and
hence translational entropy) compensates the loss of orientation
entropy (due to the nematic ordering) at sufficiently high
concentrations of rods \cite{O}. In other words, the ordering
follows as a consequence of maximizing the total entropy. More
recently it was shown that a further increase of the concentration
of TMV can also give rise to smectic ordering
\cite{DFPRL97,WMCPRL89}, where the translational invariance is
broken and the system forms a layered structure. In the 1980's and
1990's computer simulations \cite{VFPRA90,BFJCP97}, and later
density functional theories \cite{STPRL88,PHPRL88,MPRA87}, have
shown that also the smectic phase can be explained by the hard-rod
model for colloidal rods. One concludes, therefore, that the bulk
phase behavior of these systems is well understood by now.

Onsager's theory for hard rods has been extended to describe bulk
{\em mixtures} of colloidal rods. For the case of binary mixtures
of longer and shorter rods, it was found that the $IN$ transition
is accompanied by strong fractionation, such that the coexisting
nematic phase contains a relatively large fraction of the longer
rods \cite{LCHDJCP84,OLJPC85}. Later theoretical work on
long-short mixtures also showed the possibility of nematic-nematic
($N_1N_2$) demixing (driven by a peculiar competition between
orientation entropy and ideal mixing entropy), and an
isotropic-nematic-nematic ($IN_1N_2$) triple point in the phase
diagram \cite{BKP88,VLJPC93,RMPRE96,RMJCP96} of mixtures with a
length ratio more extreme than about 1:3. Later also binary
mixtures of thin and thick hard rods were considered. They were
shown to have phase diagrams similar to those of long-short
mixtures, i.e., with strong fractionation at $IN$ coexistence and
with $N_1N_2$ and $IN_1N_2$ coexistence for diameter ratios
exceeding 1:3.8 \cite{RMPRE96,RMJCP96,DRPRE97,RMDP98}, but with an
additional possibility for isotropic-isotropic ($I_1I_2$) phase
coexistence due to the depletion effect
\cite{SJJCP95,DRPRE97,RMDP98} if the diameter ratio is more
extreme than about $1:8$. Interestingly, thin-thick mixtures have
recently been realized experimentally by mixing "bare" $fd$ virus
particles (length $~1\mu$m) with ones that are "coated" with
polyethyleneglycol (PEG) \cite{PFprivate}. The diameter ratio of
these systems can be tuned by varying the ionic strength of the
solvent: due to an increasing salt concentration the effective
diameter of the (charged) bare rods shrinks because of enhanced
screening, whereas that of the PEG-coated rods is not (or hardly)
affected because of the steric nature of PEG. Exploiting this
effect allowed for the experimental study of diameter ratios up to
about 1:4.5, and $IN_1$, $IN_2$, $N_1N_2$ as well as $IN_1N_2$
triple coexistence were actually observed \cite{PFprivate}.

The present study is devoted to the planar interfaces that exist
between the coexisting bulk phases in binary mixtures of colloidal
rods. Our focus is on the calculation of both thermodynamic and
structural properties of these interfaces. The main thermodynamic
quantity of interest is the surface tension $\gamma$, and the
structural properties we will investigate are the profiles of the
density and the order parameters. It is known from the study of
the $IN$ interface of pure (one-component) suspensions of rods
that $\gamma$ depends on the angle between the interface normal
$\hat{z}$ and the director $\hat{n}$ of the nematic phase
asymptotically far from the interface \cite{PHPRA88,MM90,CN92}.
The study of Ref. \cite{CN92,CPRE93} showed that $\gamma$ is
minimal when $\hat{n} \perp \hat{z}$, and on this basis (and on
the basis of some of our own test calculations) we assume this to
be the case for mixtures as well. It is also established by now
\cite{CN92,CPRE93,SRJPCM2001} that (i) the density profile and the
nematic order parameter profile of the $IN$ interface of the pure
hard needle fluid change monotonically from their values in the
isotropic bulk phase to those in the nematic phase, (ii) the
interface thickness is of the order of the length $L$ of the rods,
and (iii) the interfacial biaxiality is small and nonmonotonic. In
this paper we will show that the density profiles in mixtures of
rods are {\em not} always monotonic, and that the interface
thickness not always of order $L$ due to the formation of
macroscopically thick wetting films close to the bulk triple
points. Some of these findings have been reported briefly
elsewhere \cite{SRPRL02}. Moreover, we will show that the tension
in mixtures of rods tends to be substantially higher than that of
the pure systems of their components.

This paper is organized as follows. In section \ref{functsect} we
introduce the Onsager functional and the basic Euler-Lagrange
equation. In section \ref{bulksect} we solve this equation for
bulk geometries, and we present a few typical  bulk phase
diagrams. In section \ref{interfsect} we present our method to
solve the Euler-Lagrange equation for interface geometries, and
study $IN_1$, $N_1N_2$ and $IN_2$ interfaces, the latter in
particular in the vicinity of the bulk $IN_1N_2$ triple point. A
summary and some discussion of our results will be presented in
section \ref{summarysect}.

\section{Density functional}
\label{functsect} We consider a fluid of hard cylinders of two
different species $\sigma=1,2$ of length $L_{\sigma}$ and diameter
$D_{\sigma}$ in a macroscopic volume $V$ at temperature $T$ and
chemical potentials $\mu_{\sigma}$. The thermodynamic properties
and the structure of the system can be determined from the grand
potential functional $\Omega[\{\rho_{\sigma}\}]$ of the
one-particle distribution functions $\rho_{\sigma}({\bf r},{\bf
\hat{\omega}})$, where ${\bf r}$ denotes the center-of-mass
coordinate of the rod of species $\sigma$ and ${\bf \hat{\omega}}$
the orientation of the long axis. The functional
$\Omega[\{\rho_{\sigma}\}]$ is such that (i) it is minimized, for
given $(\{\mu_{\sigma}\},V,T)$, by the equilibrium one-particle
distributions $\rho_{\sigma}({\bf r},{\bf \hat{\omega}})$, and
(ii) the minimal value of the functional is the equilibrium grand
potential $\Omega$ \cite{E79}.

Within the second virial approximation and in the absence of
external potentials, the functional $\Omega[\{\rho_{\sigma}\}]$
can be written \cite{O,VL} as
\begin{eqnarray}
\label{pot} \beta\Omega[\{\rho_{\sigma}\}]&=&\sum_{\sigma}\int dq
\rho_{\sigma}(q) \Big(\ln[\rho_{\sigma}(q)
L_{\sigma}^2D_{\sigma}]-1-\beta\mu_{\sigma}\Big)\nonumber\\
&&-\frac{1}{2}\sum_{\sigma{\sigma}'}\int dq dq'
f_{\sigma{\sigma}'}(q;q') \rho_{\sigma}(q)\rho_{{\sigma}'}(q'),
\end{eqnarray}
with $\beta=(kT)^{-1}$ the inverse temperature and
$f_{\sigma{\sigma}'}(q;q')$ the Mayer function of the $\sigma
\sigma'$-pair of rods with coordinates $q=\{{\bf r},{\bf
\hat{\omega}}\}$ and $q'=\{{\bf r}',{\bf \hat{\omega}}'\}$. For
hard rods, the focus of our study, $f_{\sigma{\sigma}'}(q;q')$
equals $-1$ if the rods overlap and vanishes otherwise. Onsager
argued that the second virial approximation is accurate for long
rods, and becomes even exact for isotropic and nematic bulk fluids
in the limit of vanishing diameter-to-length ratio \cite{O}. We
shall adopt this limit throughout this paper, i.e., we consider
$D_{\sigma}/L_{\sigma'} \rightarrow 0$ for any $\sigma \sigma'$
pair. Therefore, the relative shape of the rods is only
characterized by the ratios $l=L_2/L_1$ and $d=D_2/D_1$ of the
lengths and the diameters, respectively.

The minimum conditions $\delta\Omega[\{\rho_{\sigma}\}]/
\delta\rho_{\sigma}(q)=0$ on the functional lead to the set of
nonlinear integral equations
\begin{eqnarray}
\label{nonlinset} \ln[\rho_{\sigma}(q)L_{\sigma}^2D_{\sigma}]
-\sum_{{\sigma}'}\int dq' f_{\sigma{\sigma}'}(q;q')
\rho_{{\sigma}'}(q') =\beta\mu_{\sigma}
\end{eqnarray}
to be solved for the equilibrium distributions $\rho_{\sigma}(q)$.
Once determined, they can be inserted into the functional to
obtain the equilibrium value of the grand potential
\begin{eqnarray}
\label{potmin} \beta\Omega=\frac{1}{2} \sum_{\sigma}\int dq
\rho_{\sigma}(q) \Big(\ln[\rho_{\sigma}(q)L_{\sigma}^2D_{\sigma}]
-2-\beta\mu_{\sigma}\Big).
\end{eqnarray}
Note that $\Omega=-pV$ for a bulk system in a volume $V$, with
$p=p(\{\mu_{\sigma}\},T)$ the pressure. In the presence of a
planar surface or interface of area $A$ we have $\Omega=-pV+\gamma
A$ with $\gamma=\gamma(\{\mu_{\sigma}\},T)$ the surface or
interface tension.

In general, fluctuations of the interface position (capillary
waves) are important in the analysis of fluid-fluid interfaces.
The amplitude of such fluctuations is controlled by the "wetting
parameter" \cite{RW,Sh}
\begin{eqnarray}
\omega \equiv \frac{k_B T}{4\pi\gamma \xi^2},
\end{eqnarray}
where $\xi$ is the bulk correlation length. For rods of typical
length $L$ and diameter $D$ we shall see that $\xi \sim L$ and
$\gamma \sim k_B T/LD$, and hence $\omega \sim D/L$, i.e.,
$\omega$ vanishes in the Onsager limit. As a consequence the
capillary-wave fluctuations are unimportant, i.e., the mean-field
density functional (\ref{pot}) is sufficient to describe
interfacial phenomena in fluids of long hard rods.

\section{Bulk phase diagrams}
\label{bulksect} The bulk thermodynamic properties of binary
hard-rod fluids were studied extensively within Onsager theory.
The minimization of the functional was either performed
variationally \cite{BKP88,VLJPC93,SJJCP95,RMPRE96, RMJCP96,HMP99}
or through a fully numerical solution
\cite{LCHDJCP84,RMDP98,SRJPCM2001}. We adopted the latter approach
as it can be easily generalized for inhomogeneous systems. For
clarity we briefly repeat the essential points of the method and
summarize the available results.

The bulk distribution functions of the isotropic and nematic phase
are translationally invariant, i.e., $\rho_{\sigma}({\bf
r},\hat{\omega})=\rho_{\sigma}(\hat{\omega})$, which allows to
reduce Eq. (\ref{nonlinset}) to
\begin{equation}
\label{stathom}
\ln[\rho_{\sigma}(\hat{\omega})L_{\sigma}^2D_{\sigma}]+\sum_{\sigma'}
\int d\hat{\omega}'E_{\sigma{\sigma}'}(\hat{\omega},\hat{\omega}')
\rho_{{\sigma}'}(\hat{\omega}') =\beta\mu_{\sigma},
\end{equation}
with $E_{\sigma{\sigma}'}$ the excluded volume of a pair of
cylinders of species $\sigma$ and $\sigma'$ given by \cite{VL}
\begin{eqnarray}
\label{exclvol} E_{\sigma{\sigma}'}(\hat{\omega},\hat{\omega}')&=&
-\int d{\bf r}'f_{\sigma{\sigma}'}({\bf r},\hat{\omega};{\bf
r}',\hat{\omega}')\nonumber\\
&=&L_{\sigma}L_{\sigma'}(D_{\sigma}+D_{\sigma'})|\sin\gamma|
\end{eqnarray}
in terms of the angle $\gamma$ between $\hat{\omega}$ and
$\hat{\omega}'$, i.e., $\gamma=\arccos (\hat{\omega} \cdot
\hat{\omega}')$. Note that additional $O(LD^2)$ terms are being
ignored in Eq. (\ref{stathom}), in line with the needle limit
($D_{\sigma}/L_{\sigma'} \rightarrow 0$) of interest here.

At sufficiently low $\{ \beta\mu_{\sigma} \}$ the only stable
solution of Eq. (\ref{stathom}) is the isotropic distribution
$\rho_{\sigma}^{I}(\hat{\omega})=n_{\sigma}/{(4\pi)}$, with
$n_{\sigma}=\int d\hat{\omega}\rho_{\sigma}(\hat{\omega})$ the
bulk number density of species $\sigma$. As $\mu_{\sigma}$ are
high enough, one or, possibly, two sets of stable uniaxial
solutions $\rho_{\sigma}^{N}(\hat{\omega})= \rho_{\sigma}(\theta)$
exist, with $\theta=\arccos(\hat{\omega} \cdot\hat{n})$ the angle
between $\hat{\omega}$ and the nematic director $\hat{n}$. These
distributions have "up-down" symmetry,
$\rho_{\sigma}(\theta)=\rho_{\sigma}(\pi-\theta)$, hence
$\rho_{\sigma}(\theta)$ needs only to be determined for
$\theta\in[0,\pi/2]$. Using an equidistant $\theta$-grid of
$N_{\theta}=30$ points $\theta_i\in[0,\pi/2]$, where $1\leq i\leq
N_{\theta}$, we iteratively solve Eq. (\ref{stathom}) for the set
of $2N_{\theta}$ equations in order to find
$\rho_{\sigma}(\theta_i)$ numerically. The integral in Eq.
(\ref{stathom}) is calculated with the trapezoidal rule.
Coexistence of different phases $\{I,N_1,N_2\}$ can be determined
by imposing conditions of mechanical and chemical equilibrium.

In order to gauge the accuracy of the chosen $\theta$-grid we
calculate the resulting densities of the coexisting isotropic and
nematic phase of the one-component system. We find $n^I L^2D
(\pi/4)=3.281 \pm 0.001$, $n^NL^2D(\pi/4)=4.172 \pm 0.001$; the
nematic order parameters of the two coexisting phases are
$S^I=0.008 \pm 0.001$ and $S^N=0.791 \pm 0.001$, and the pressure
$(\pi/4)\beta pL^2D=14.045 \pm 0.001$. These data, based on
$N_{\theta}=30$, differ by less then a percent from the most
accurate results available in the literature \cite{VL}, which we
can reproduce with $N_{\theta} \geq 80$. In order to have
consistency between bulk and interfacial results we take
$N_{\theta}=30$ in most of our calculations. The exception is the
case of long-short mixtures in the nematic phase, which requires
$N_{\theta}=50$ for acceptable accuracy.

In Fig. \ref{thinthickphase} we show both pressure-composition (a)
and density-density (b) representations of bulk phase diagrams of
thin-thick binary mixtures ($L_{\sigma}=L,D_2>D_1$) for several
diameter ratios $d\equiv D_2/D_1$. In Fig. \ref{thinthickphase}(a)
the composition variable $x=n_2/(n_1+n_2)$ denotes the mole
fraction of thick rods and $f=(p-p_{thick})/(p_{thin}-p_{thick})$
is a dimensionless shifted pressure which takes the values $f=1,0$
at isotropic-nematic coexistence of the pure-thin ($x=0$) and
pure-thick ($x=1$) systems, respectively. Note that $(\pi /4)
\beta p_{thin}L^2D_1= (\pi /4) \beta p_{thick}L^2D_2=14.045$,
i.e., $p_{thick}=p_{thin}/d$, and that the tie-lines connecting
coexisting phases are horizontal in the $f-x$ representation of
Fig. \ref{thinthickphase}(a).

At low pressures (or low densities) the phase diagrams show an
isotropic ($I$) phase and at higher pressures (or densities) one
($d=3.0,3.5$) or two ($d=4.0,4.2$) nematic phases ($N_1$ and
$N_2$). For diameter ratios $d=3.0,3.5$ the phase diagram is
spindlelike, and the only feature is a strong fractionation at
coexistence, such that the nematic phase is relatively rich in
thick rods and the isotropic phase in thin ones. The reason behind
this fractionation is the relatively large excluded volume in
interactions of the thick rods, which makes them more susceptible
to orientational ordering \cite{LCHDJCP84,RMJCP96,RMDP98}. The
fractionation of isotropic-nematic coexistence becomes stronger
for increasing $d$. For $3.8<d<4.29$ the bulk phase diagram
develops nematic-nematic ($N_1N_2$) coexistence in a pressure
regime $p_t<p<p_c$, with $p_t$ the triple-point pressure and $p_c$
the critical (consolute) pressure. For $d=4.0$ the consolute point
is indicated by ($\ast$) in Fig. \ref{thinthickphase}. The
mechanism of this demixing transition was spelled out in detail in
Refs.\cite{VLJPC93,RMDP98}, and involves a competition between
orientational entropy (favoring demixing) and entropy of mixing
(favoring mixing). Interestingly, the width of fractionation gap
$\Delta x=x_{N_1}-x_{N_2}$ for the triple point $N_1$ and $N_2$
phases scales linearly with the triple-point pressure $p_t$. The
critical pressure of the $N_1N_2$ transition diverges as
$d\rightarrow 4.29$ \cite{VLJPC93,RMDP98}. For $d>3.8$ the lower
bound of the $N_1N_2$ coexistence is an $IN_1N_2$ triple point,
indicated by triangles ($\bigtriangleup$) in Fig.
\ref{thinthickphase} for $d=4.0$. With increasing $d$ the triple
point $I$ and $N_1$ phases approach the pure-thin bulk coexistence
(i.e., $x_{I,N_1} \rightarrow 0$), whereas the composition of the
triple point $N_2$ phase shifts to a pure-thick phase ($x_{N_2}
\rightarrow 1$).

The $f-x$ representation is convenient for our analysis, whereas
the densities (volume fractions) of thin and thick rods are
experimental control parameters \cite{SF}. For this reason the
same phase diagrams of thin-thick binary mixtures are shown in
Fig. \ref{thinthickphase}(b) in density-density representation,
with $n_1^{*}=n_1LD_1^2(\pi/4)$ and $n_2^{*}=n_1LD_2^2(\pi/4)$
being the dimensionless bulk number densities of thin and thick
rods, respectively. In this representation the tie-lines are no
longer horizontal.

\begin{figure}[h]\centering
\includegraphics[width=8.5cm]{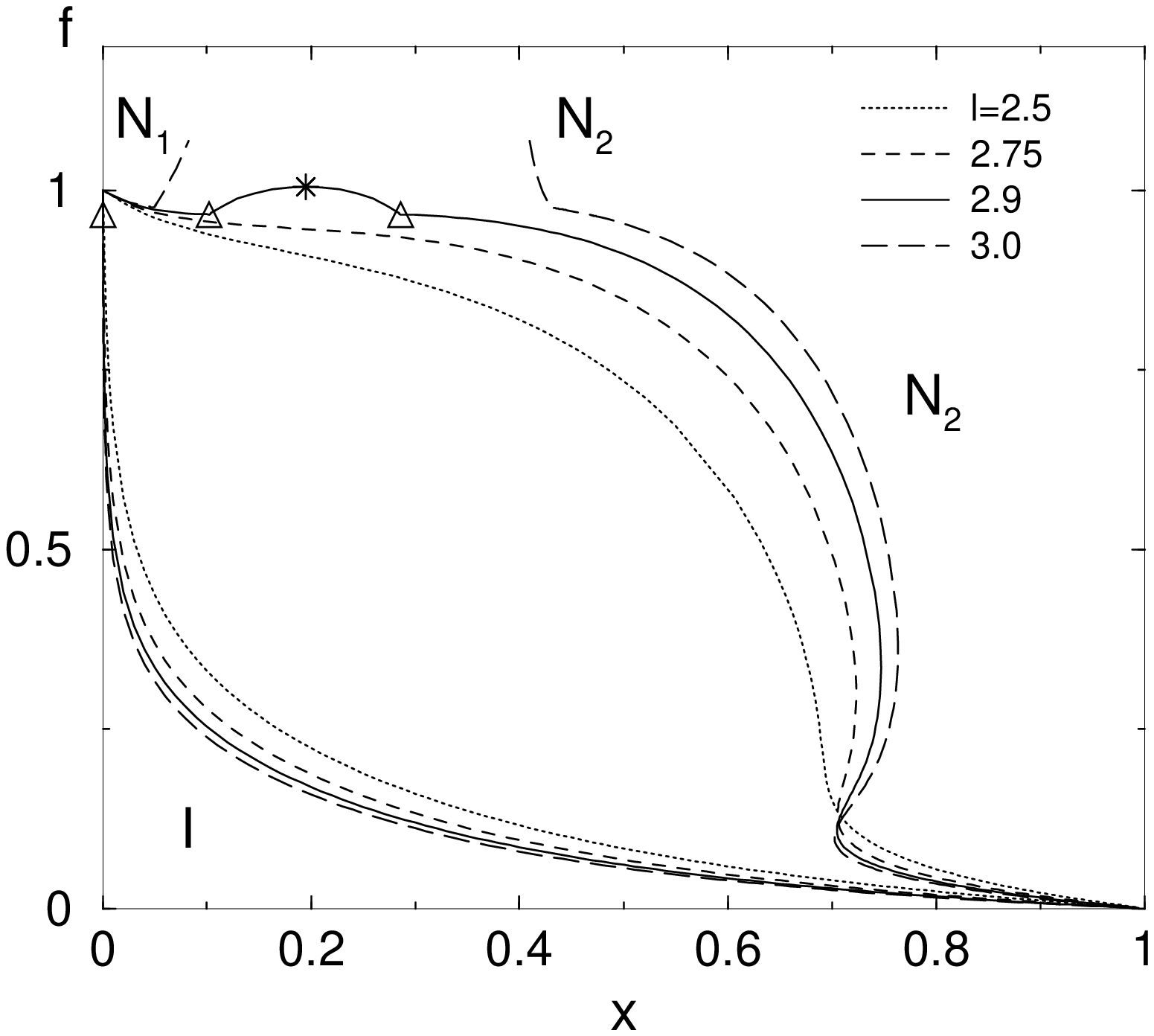}
\caption{\label{longshortphase} K. Shundyak and R. van Roij}
\end{figure}

In Fig. \ref{longshortphase} a set of bulk phase diagrams for
long-short binary mixtures ($D_{\sigma}=D, L_2>L_1$) for several
length ratios $l\equiv L_2/L_1$ is presented. Here
$x=n_2/(n_1+n_2)$ denotes the fraction of long rods and
$f=(p-p_{long})/(p_{short}-p_{long})$ with
$p_{long}=p_{short}/l^2$. All characteristic features of the phase
diagrams are the same as in thin-thick mixtures. The fractionation
of the coexisting $IN_2$ and $N_1N_2$ phases has a strong
dependence on $l$, and limits the values accessible for
calculations to $l \leq 3.1$ for the chosen grids and the required
accuracy. The main reason is that the nematic ordering in the
triple point $N_2$ phase is very pronounced, requiring a fine grid
\cite{RMDP98}.

\section{Free interfaces}
\label{interfsect} We now turn to the thermodynamics and the
structure of the free interfaces between the coexisting phases. We
assume that the interfaces are planar, with surface normal
$\hat{z}$. The nematic director $\hat{n}$ of the asymptotic
nematic bulk phase(s) can, in general, have a nontrivial tilt
angle $\theta_{t}=\arccos(\hat{n} \cdot\hat{z})$ with respect to
the interface normal. In the present calculations we restrict
attention to $\theta_{t}=\pi/2$, i.e., $\hat{n} \perp \hat{z}$.
This geometry is known to be thermodynamically favorable because
of its minimal surface tension \cite{CN92,MASPRE01}.

The equilibrium distribution functions
$\rho_{\sigma}(z,\hat{\omega})$ depend on the spatial coordinate
$z=\hat{z}\cdot \bf{r}$, and angular coordinates
$\hat{\omega}=(\theta,\varphi)$ defined by
$\cos\theta=\hat{n}\cdot\hat{\omega}$ and
$\sin\theta\sin\varphi=\hat{z}\cdot\hat{\omega}$. These functions
are solutions of the Euler-Lagrange equations Eq.
(\ref{nonlinset}) at the coexisting chemical potentials $\beta
\mu_{\sigma}= \beta \mu_{\sigma}^{coex}$, with boundary conditions
$\rho_{\sigma}(z\rightarrow\pm\infty,\hat{\omega})=
\rho_{\sigma}^{(\pm)}(\theta)$ being the two coexisting bulk
distributions (labelled by $(+)$ and $(-)$ here for brevity).

The planar symmetry, i.e., the independence of
$\rho_{\sigma}(z,\hat{\omega})$ of the in-plane coordinates $x$
and $y$, allows for a reduction of the numerics, since the
"excluded slab" $\mathcal{
K}_{\sigma{\sigma}'}(z,\hat{\omega},z',{\hat{\omega}}')=-\int
dx'dy' f_{\sigma{\sigma}'}(\mathbf{r},\hat{\omega};
\mathbf{r}',\hat{\omega}')$ can be calculated analytically
\cite{PHPRA88,SRJPCM2001}. This reduces the Euler-Lagrange
equations to
\begin{eqnarray}
\label{rhoreduced} \beta\mu_{\sigma}^{coex}&=&\ln[\rho_{\sigma}
(z,\hat{\omega})L_{\sigma}^2D_{\sigma}]+ \sum_{{\sigma}'}
\int d z'\int d \hat{\omega}' \nonumber\\
&&\times \mathcal{K}_{\sigma{\sigma}'}(|z-z'|,\hat{\omega},
\hat{\omega}') \rho_{{\sigma}'}(z',\hat{\omega}'),
\end{eqnarray}
where the expression for
$\mathcal{K}_{\sigma{\sigma}'}(|z|,\hat{\omega},\hat{\omega}')$ is
given in Appendix.

In principle one could now solve Eq. (\ref{rhoreduced}) on a
($z,\theta,\varphi$) grid. However, the numerical efforts can be
further reduced if one realizes that biaxiality, i.e., the
$\varphi$-dependence, is weak \cite{CPRE93,SRJPCM2001}. In that
case the truncated expansion
\begin{eqnarray}
\label{expansrho}\rho_{\sigma} (z,\theta,\varphi)&=&\sum_{m=0}^M
\rho_{\sigma,m} (z,\theta) \cos(2m\varphi)
\end{eqnarray}
is expected to be accurate for small $M$, and hence only a few
"coefficients" $\rho_{\sigma,m} (z,\theta)$ ($m \leq M$) need to
be determined on a ($z,\theta$)-grid. It is important to realize,
however, that Eq. (\ref{expansrho}) implicitly assumes that the
nematic director $\hat{n}$ does not vary in space. The
coefficients $\rho_{\sigma,m} (z,\theta)$ follow from an insertion
of Eq. (\ref{expansrho}) into Eq. (\ref{rhoreduced}),
multiplication by $\cos(2m\varphi)$, and integration over
$\varphi$ ($0 \leq \varphi \leq 2 \pi $). For $M=0$ this yields
\begin{eqnarray}
\label{rho0}
\beta\mu_{\sigma}^{coex}&=&\ln[\rho_{\sigma,0}(z,\theta)
L_{\sigma}^2D_{\sigma}]+\sum_{{\sigma}'}\int dz' d\theta'
\sin\theta' \nonumber\\
&&\times \mathcal{K}_{\sigma{\sigma}'}^{00}(z-z',\theta,\theta')
\rho_{\sigma',0}(z',\theta'),\nonumber\\
\end{eqnarray}
where
$\mathcal{K}_{\sigma{\sigma}'}^{00}(z-z',\theta,\theta')=(2\pi)^{-1}\int_0^{2\pi}
\int_0^{2\pi} d\varphi d\varphi' \mathcal{K}_{\sigma{\sigma}'}
(z-z',\hat{\omega},\hat{\omega}')$ is the doubly azimuthally
integrated excluded slab, which we determine numerically once on
an appropriate grid.

The lowest-order correction that takes into account biaxiality
results from $M=1$, and yields $\rho_{\sigma,m} (z,\theta)$,
($m=0,1$) from the coupled set of equations
\begin{widetext}
\begin{eqnarray}
\beta\mu_{\sigma}^{coex}&=&\ln[\rho_{\sigma,0}(z,\theta)L_{\sigma}^2D_{\sigma}]+
I_0 \biggl( \frac{\rho_{\sigma,1}(z,\theta)}
{\rho_{\sigma,0}(z,\theta)} \biggr) \nonumber\\
&&+\sum_{{\sigma}'}\int dz' d\theta' \sin\theta' \Bigl(
\mathcal{K}_{\sigma{\sigma}'}^{00}(z-z',\theta,\theta')\rho_{\sigma',0}(z',\theta')
+\mathcal{K}_{\sigma{\sigma}'}^{01}(z-z',\theta,\theta')\rho_{\sigma',1}(z',\theta') \Bigr),\nonumber\\
0&=&I_1 \biggl( \frac{\rho_{\sigma,1}(z,\theta)}
{\rho_{\sigma,0}(z,\theta)} \biggr)+\sum_{{\sigma}'}\int dz'
d\theta' \sin\theta' \Big(
\mathcal{K}_{\sigma{\sigma}'}^{10}(z-z',\theta,\theta')\rho_{\sigma',0}(z',\theta')\nonumber\\
&&+\mathcal{K}_{\sigma{\sigma}'}^{11}(z-z',\theta,\theta')
\rho_{\sigma',1}(z',\theta') \Big), \label{rho1}
\end{eqnarray}
with
$\mathcal{K}_{\sigma{\sigma}'}^{km}(z-z',\theta,\theta')=(2\pi)^{-1}\int_0^{2\pi}
\int_0^{2\pi} d\varphi d\varphi' \cos(2k\varphi) \cos(2m\varphi')
\mathcal{K}_{\sigma{\sigma}'} (z-z',\hat{\omega},\hat{\omega}')$,
$k,m=\{0,1\}$, again to be determined numerically only once, with
\begin{eqnarray*}
I_0(x)&=&\frac{1}{2\pi}\int_0^{2\pi}d\varphi \ln\big(1+x
\cos(2\varphi)\big)=\ln\frac{1+\sqrt{1-x^2}}{2},\nonumber\\
I_1(x)&=&\frac{1}{2\pi}\int_0^{2\pi}d\varphi \cos(2\varphi)
\ln\big(1+x \cos(2\varphi)\big)= \frac{1-\sqrt{1-x^2}}{x},
\label{i0i1}
\end{eqnarray*}
for $|x|<1$.
\end{widetext}

Note that the boundary conditions imply that
$\rho_{\sigma,1}(z,\theta)\rightarrow 0$ for
$|z|\rightarrow\infty$. In general, the solutions
$\rho_{\sigma,0}(z,\theta)$ for $M=0$ are not identical to
$\rho_{\sigma,0}(z,\theta)$ for $M=1$, but the difference is small
in most cases since $|\rho_{\sigma,1}(z,\theta)L_\sigma
D_\sigma^2|\ll 1$. In the remainder of this paper we shall mainly
concentrate on solutions of Eq. (\ref{rho0}), although some
results of Eqs. (\ref{rho1}) will be discussed.

By iteration of Eq. (\ref{rho0}) (or Eqs. (\ref{rho1})) with the
appropriate boundary conditions we calculated
$\rho_{\sigma,0}(z,\theta)$ (and $\rho_{\sigma,1}(z,\theta)$) for
a number of state points $\mu_{\sigma}^{coex}$ on the $IN_1$,
$IN_2$ and $N_1N_2$ binodals. We used an equidistant spatial grid
of $N_z=200$ points $z_i\in[-5L,5L]$, an equidistant angular grid
of $N_{\theta}=30$ points ${\theta}_j\in[0,\pi/2]$ for thin-thick
mixtures or an angular grid of $N_{\theta}=50$ points
${\theta}_j\in[0,\pi/2]$ for long-short mixtures. From the
equilibrium distributions $\rho_{\sigma,0}(z,\theta)$ we
calculated the local density and the nematic order parameter
profiles
\begin{eqnarray*}
n_{\sigma}(z)&=&4\pi \int_{0}^{\pi/2} d \theta \sin\theta
\rho_{\sigma,0}(z,\theta)\\
S_{\sigma}(z)&=&4\pi \int_{0}^{\pi/2} d \theta \sin\theta
P_2(\cos\theta) \rho_{\sigma,0}(z,\theta)/n_{\sigma}(z),
\end{eqnarray*}
with $P_2(x)=(3x^2-1)/2$ the second Legendre polynomial. In the
case of iterating Eqs. (\ref{rho1}) the biaxiality is defined as
\cite{CPRE93,SRJPCM2001}
\begin{eqnarray*}
\Delta_{\sigma}(z)&=& \left\langle \frac{3}{2} \sin^2\theta\cos
2\varphi \right\rangle_{\sigma}\\
&=&\frac{3}{4n_{\sigma}(z)} \int_0^{\pi/2} d\theta\sin^3\theta
\rho_{\sigma,1}(z,\theta).
\end{eqnarray*}

The interface thickness $t$ is defined as $t=|z_+-z_-|$ where
$z_\pm$ are solutions of ${n_1}'''(z)=0$, where a prime denotes
differentiation with respect to $z$. As this equation has a set of
solutions in every interfacial region, we choose $(z_\pm)$ be the
outermost ones, i.e., the nearest to the bulk phases. This
criterion provides a single measure for the thickness of both
monotonic and non-monotonic profiles, with and without a thick
film in between the asymptotic bulk phases at $z\rightarrow \pm
\infty$. Also thin (or short) rods have a smaller excluded volume
and a non-vanishing concentration in both coexisting phases, so
their density is a convenient representation of structural changes
within the interface. The interfacial width for the one-component
$IN$ interface is, with the present definition, given by
$t/L=0.697$. We have checked that other definitions of the
thickness lead to similar results.

\subsection{$IN_1$ and $N_1N_2$ interfaces.}
The $IN_1$ interfaces exist only in a small pressure regime
$p_{thin}\geq p\geq p_t$. They closely resemble the $IN$ interface
of the pure hard-rod fluid, i.e., the profiles of the order
parameters $S_{\sigma}(z)$ and the densities $n_{\sigma}(z)$
change monotonically from the bulk values in the $I$ phase to
those in the $N_1$ phase.

\begin{figure}[h]\centering
\includegraphics[width=8.5cm]{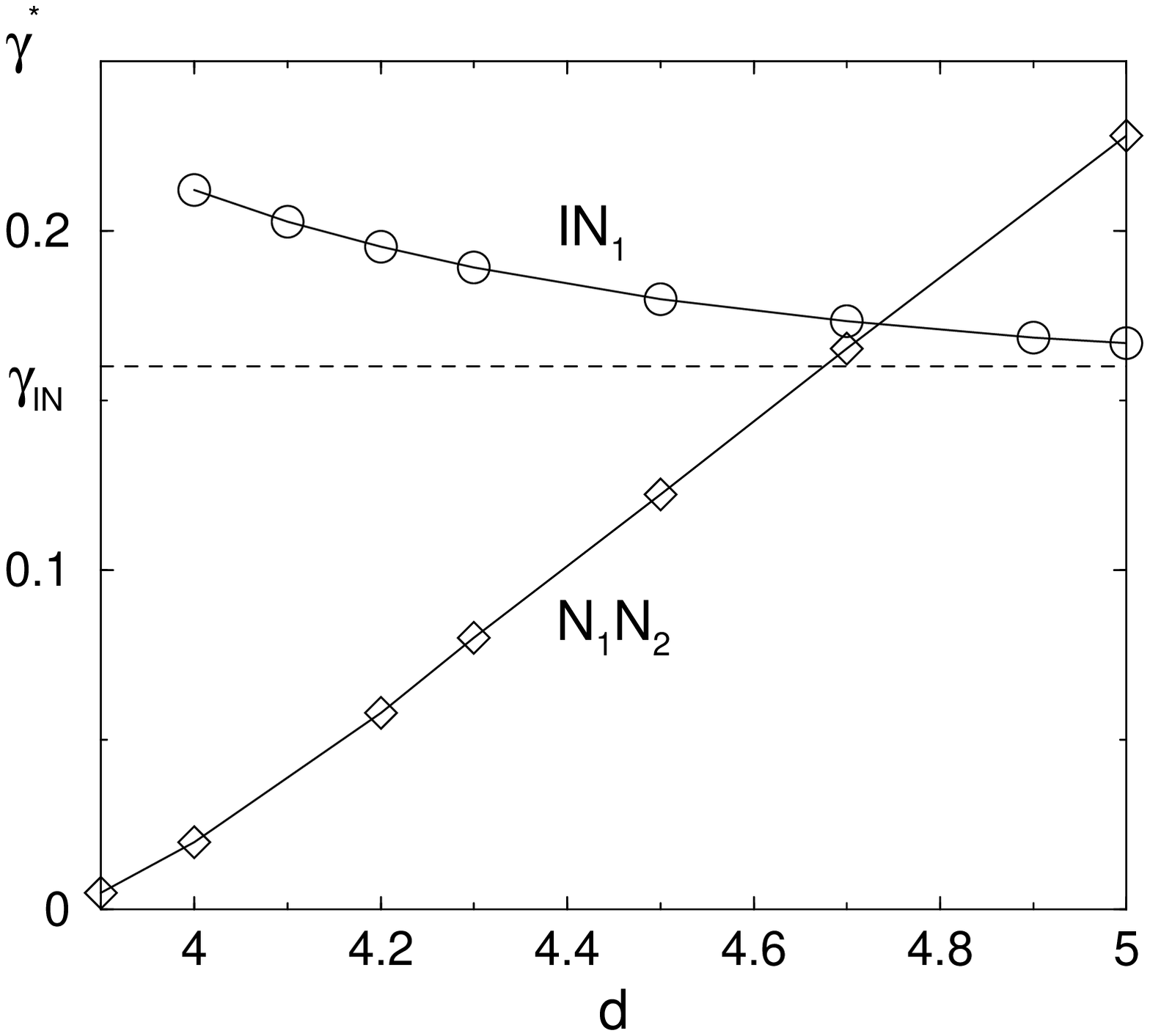}
\caption{\label{in1n2surf} K. Shundyak and R. van Roij}
\end{figure}

The thickness of the $IN_1$ interface is of order $L$ which is
similar to that of the pure system. With increasing $d$ the $IN_1$
surface tension at triple phase coexistence ($p=p_t$) decreases
monotonically to the $IN$ surface tension of the pure system, as
shown in Fig. \ref{in1n2surf}, where the dimensionless surface
tension $\gamma^{*}=\beta \gamma /LD_1$ is plotted. This is
expected from an inspection of the phase diagrams as $x_{I,N_1}
\rightarrow 0$ for increasing $d$. For the diameter ratio $d=4.0$
the surface tension at $p=p_t$ is given by
$\gamma_{IN_1}^{*}=0.209\pm 0.001$ with the same level of accuracy
for all other calculations of the surface tensions. For the
long-short mixtures the behavior of the surface tension at the
$IN_1$ interface as a function of length ratio $l$ is very
similar. For $l=3.0$ the surface tension at $p=p_t$ is given by
$\gamma_{IN_1}^{*}=0.212 \pm 0.001$.

In general, the order parameter and density profiles are shifted
with respect to each other. Such a shift can be characterized by
the distance $\delta=|z_n-z_S|$ between the centers $z_n$ and
$z_S$ of the density and order parameter profiles, defined by
\cite{CN92}
\begin{eqnarray*}
n(z_n)&=&n_{-}+\frac{1}{2} (n_{+}-n_{-}),\\
S(z_S)&=&S_{-}+\frac{1}{2}(S_{+}-S_{-}),
\end{eqnarray*}
where $+/-$ indicate asymptotic bulk values. A non-zero shift
$\delta$ reflects the fact that the thickness of the interface is
different for rods of different orientations. For monodisperse
rods it was found that $\delta=0.50L$ \cite{CN92}. For binary
mixtures $\delta$ can be determined for each component separately.
For thin-thick mixtures with $d=4.0$ the $IN_1$ interface at
triple phase coexistence shows $\delta_{thin}=0.35L$ and
$\delta_{thick} =0.55L$. For long-short mixtures with $l=3.0$ the
effect is similar for the short rods ($\delta_{short}=0.37L_1$)
and much more pronounced for the long rods
($\delta_{long}=1.54L_1$, i.e., $\delta_{long} \simeq 0.51L_2$).

The profiles of $S_{\sigma}(z)$ and $n_{\sigma}(z)$ at $N_1N_2$
interfaces are also monotonic. For the diameter ratio $d=4.0$ at
$p=p_t$ the interface thickness is given by $t/L=0.592$ and the
surface tension $\gamma_{N_1N_2}^{*}=0.019\pm 0.001$, which is an
order of magnitude smaller than $\gamma_{IN_1}^{*}$. Upon the
approach of the critical point $t/L \rightarrow \infty$ and the
surface tension vanishes. For $d>4.0$ surface tension
$\gamma_{N_1N_2}^{*}$ (at $p=p_t$) increases approximately
linearly with $d$ as shown in Fig. \ref{in1n2surf}.

\begin{figure}[h]\centering
\includegraphics[width=8.5cm]{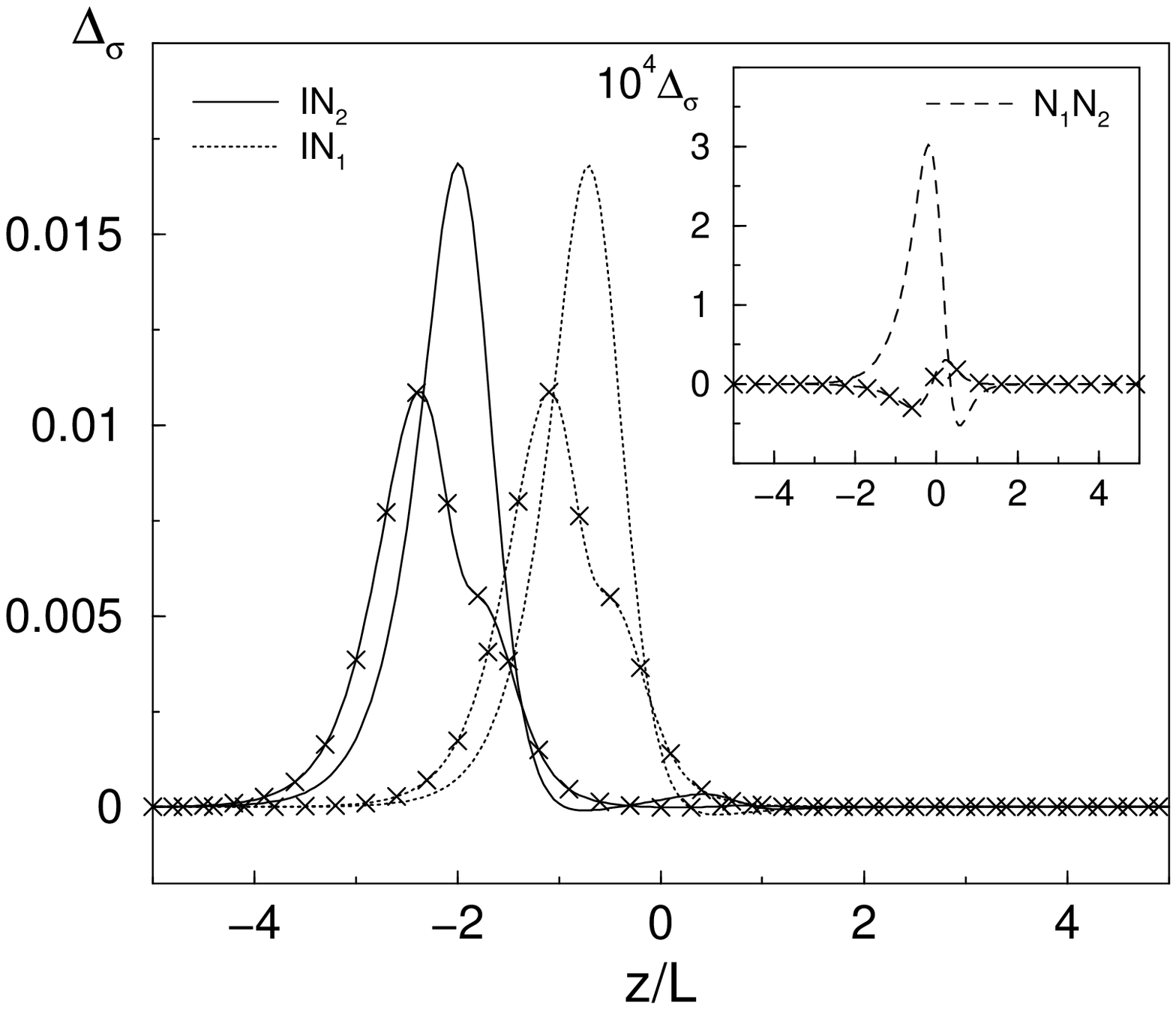}
\caption{\label{biax} K. Shundyak and R. van Roij}
\end{figure}

The biaxiality is found to be small in both the $IN_1$ and the
$N_1N_2$ interfaces. In Fig. \ref{biax} we present the profiles
$\Delta_{\sigma}(z)$ of the triple point $IN_1$ and $N_1N_2$
(inset) interfaces of the thin-thick mixture with $d=4.0$, as well
as that of the $IN_2$ interface to be discussed later. The marked
curves represent $\Delta_2(z)$ (thick rods), the unmarked ones
$\Delta_1(z)$ (thin rods). Figure \ref{biax} reveals that
$|\Delta_{\sigma}(z)|<0.017$ in the $IN_1$ interface, and
$|\Delta_{\sigma}(z)|<4.0\cdot 10^{-4}$ in the $N_1N_2$ interface.
Such small biaxialities indicate that the expansion of Eq.
\ref{expansrho}, truncated at $M=1$, is accurate for calculating
$\Delta_{\sigma}(z)$, while a truncation at $M=0$ yields accurate
tensions and density profiles. In fact, we checked that the
difference between the tensions based on uniaxial ($M=0$) and
biaxial ($M=1$) profiles falls within the numerical accuracy,
i.e., less than $1\%$. Our numerical data for the $IN_1$ interface
is consistent with that of Ref. \cite{CPRE93,SRJPCM2001}. Even
though the magnitude of $\Delta_{\sigma}(z)$ is small, it is
interesting to consider the structure of the profiles in some more
detail. The first observation we make is that
$\Delta_{\sigma}(z)>0 \ (<0)$ at the isotropic (nematic) side of
the $IN_1$ interface for both species $\sigma=1,2$. This indicates
that rods at the $I$-side of the interface have a (small)
preference for splay in the $XY$-plane, whereas those at the $N_1$
side tend to "stick" through the interface (into the $I$ side). A
similar effect exists in the $N_1N_2$ interface, but now both
species have an opposite tendency (see inset): the thin rods splay
in the $XY$ plane at the $N_1$ side, whereas the thick ones
"stick" through, and vice versa at the $N_2$ side. Recall,
however, that these effects are small.

\subsection{The $IN_2$ interfaces.}
\begin{figure}[h]\centering
\includegraphics[width=8.5cm]{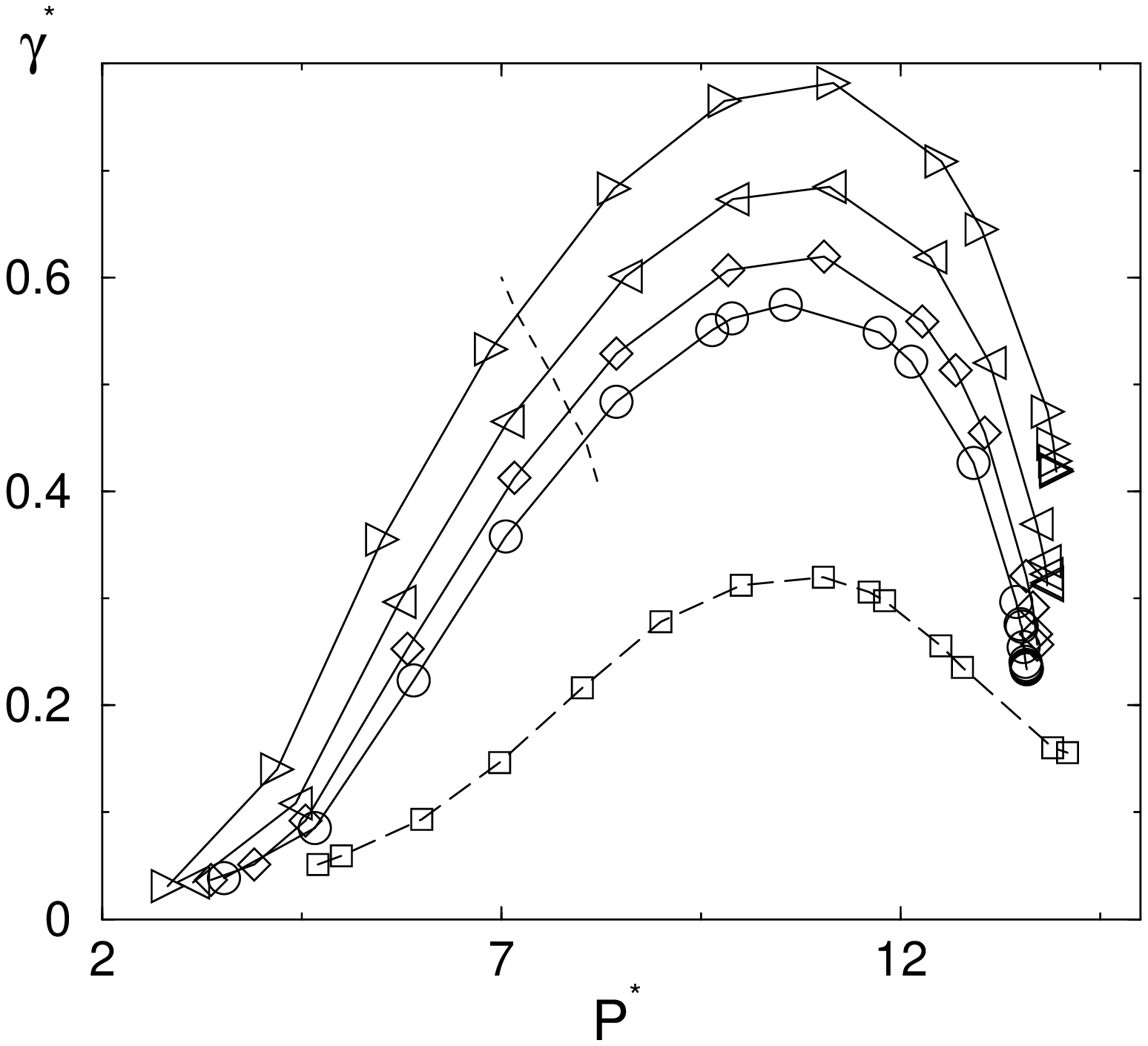}
\caption{\label{in2surf} K. Shundyak and R. van Roij}
\end{figure}

The $IN_2$ interfaces exist, for $d>3.8$, in a pressure regime
$p_{thick}<p<p_t$. The properties of the $IN_2$ interfaces depend
strongly on the pressure difference with the triple-point
($IN_1N_2$ phase coexistence). The surface tension of the $IN_2$
interface shows a non-monotonic dependence on the bulk pressure
$p$. It develops a maximum, which is several times larger than a
linear interpolation between the tension of the two pure systems,
as shown in Fig. \ref{in2surf}. It turns out that the
non-monotonic character of $\gamma_{IN_2}(p)$ is related to the
fractionation at the $IN_2$ coexistence, i.e., a larger
composition change through the interface leads to a larger
interfacial tension. However, the surface tension which
corresponds to the pressure of maximal bulk fractionation
(indicated by the dashed line in Fig. \ref{in2surf}) is lower than
the maximum of $\gamma_{IN_2}(p)$. The maximal interface stiffness
grows with species diameter ratio $d$ as the composition
difference between $I$ and $N_2$ phases increases (see Fig.
\ref{thinthickphase}). We have also compared the maximal surface
tensions for different orientations of the director ($\hat{n}
\perp \hat{z}$ and $\hat{n}
\parallel \hat{z}$) in several thin-thick mixtures and found
the geometry $\hat{n} \perp \hat{z}$ to be thermodynamically
stable, i.e., $\gamma_{\hat{n} \perp \hat{z}}<\gamma_{\hat{n}
\parallel \hat{z}}$ by at least a factor of two.

The relatively large surface tension of a mixture of rods compared
to that of the pure systems of its components may well be an
explanation for the relatively large tensions that were measured
in suspensions of cellulose \cite{CGL02}, which are known to be
very polydisperse. This remains to be investigated in detail,
however.

\begin{figure}[h]\centering
\includegraphics[width=8.5cm]{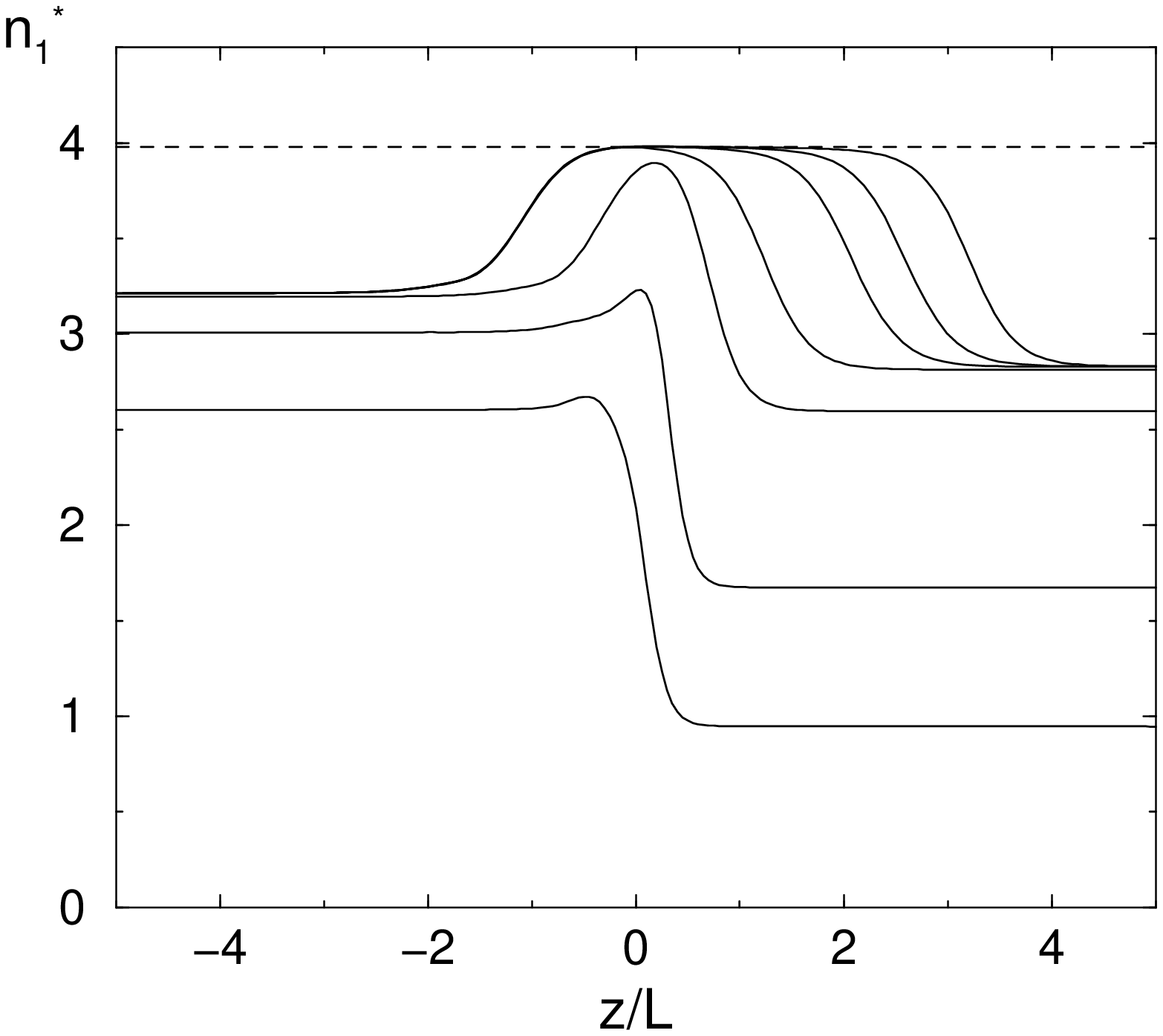}
\includegraphics[width=8.5cm]{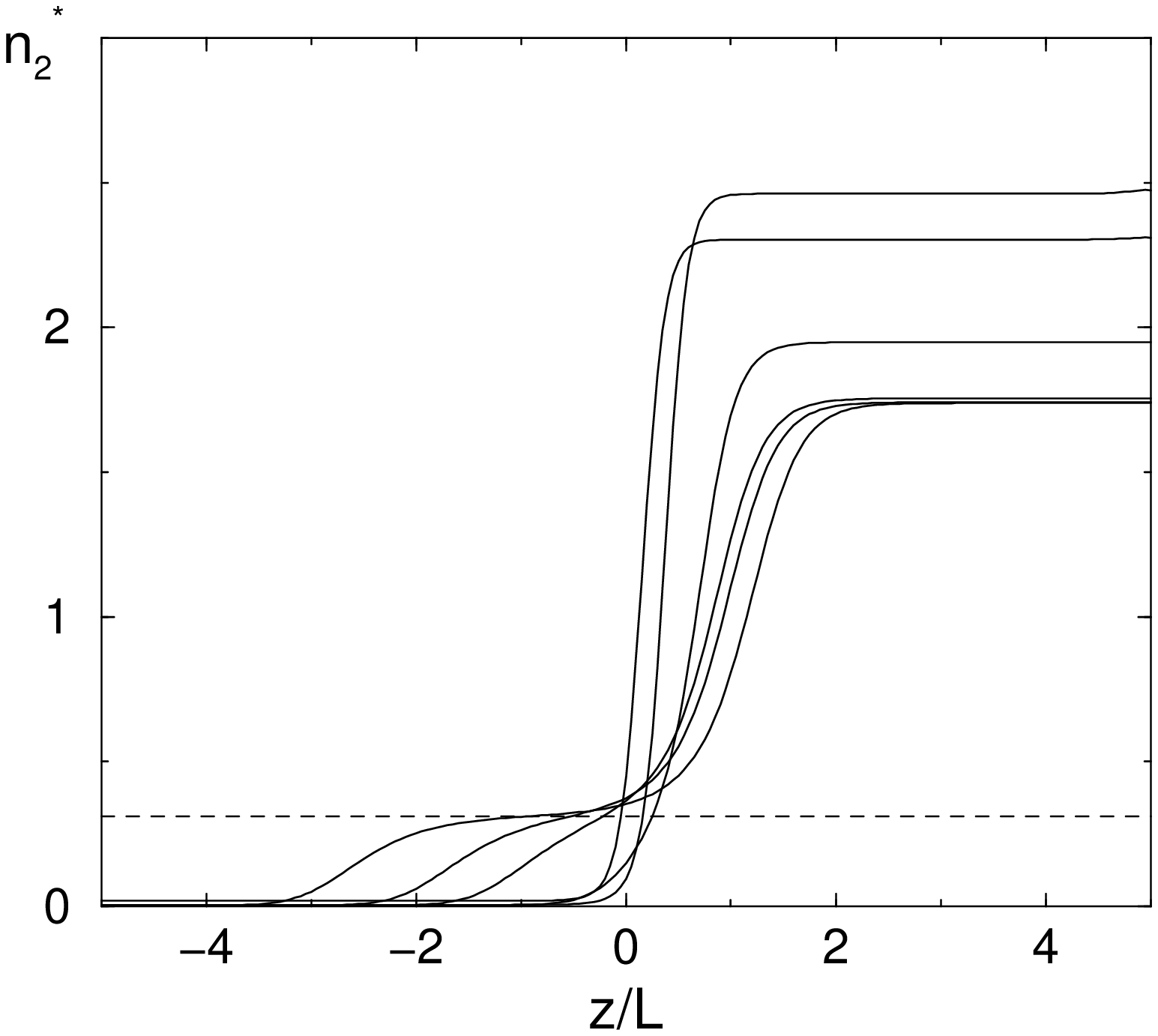}
\caption{\label{in2densn1} K. Shundyak and R. van Roij}
\end{figure}

The dimensionless undersaturation $\epsilon=1-p/p_t$ is a
convenient measure of the pressure difference with the triple
point. For $0.01<\epsilon<0.5$ the profiles of the order
parameters $S_{\sigma}(z)$ and the density of the thick component
$n_2(z)$ are smooth and monotonic, whereas $n_1(z)$ shows an
accumulation of thin rods at the isotropic side of the interface.
This effect becomes more pronounced for small undersaturation,
i.e., $\epsilon \rightarrow 0$, when a film of the nematic phase
$N_1$ appears in the $IN_2$ interface. Note that the $N_1$ phase
is a metastable bulk phase for any $\epsilon >0$, so the film
thickness is finite. For $d=4.0$ several profiles $n_1(z)$ and
$n_2(z)$ for different values of $\epsilon$ are presented in Fig.
\ref{in2densn1}, which clearly shows the film formation when
$\epsilon \rightarrow 0$. The asymptotic densities at
$z\rightarrow \pm \infty$ are those of the coexisting $I$ and
$N_2$ bulk phases (at the corresponding $\epsilon$). Using
translational invariance of the interface between the bulk phases,
we have shifted the profiles with respect to each other such that
their $IN_1$ interfaces coincide. This shows that the local
density of thin (thick) rods in the growing film remains constant,
and exactly corresponds to the thin (thick) -rod density of the
bulk triple point phase $N_1$ (indicated by the dashed lines in
Fig. \ref{in2densn1}). The same identification can be made for all
$d$ (or $l$ for long-short mixtures) as well as for the order
parameter profiles $S_{\sigma}(z)$.

The biaxiality of the $IN_2$ interface was found to be small. A
typical profile for thin-thick binary mixture with $d=4.0$ at
$\epsilon=5\cdot 10^{-4}$ is presented in Fig. \ref{biax}. The
$IN_2$ biaxiality profile can be considered as a composition of
the (earlier discussed) $IN_1$ and $N_1N_2$ profiles which is
expected as the thickness of the wetting $N_1$ film is larger than
$L$.

\begin{figure}[h]\centering
\includegraphics[width=8.5cm]{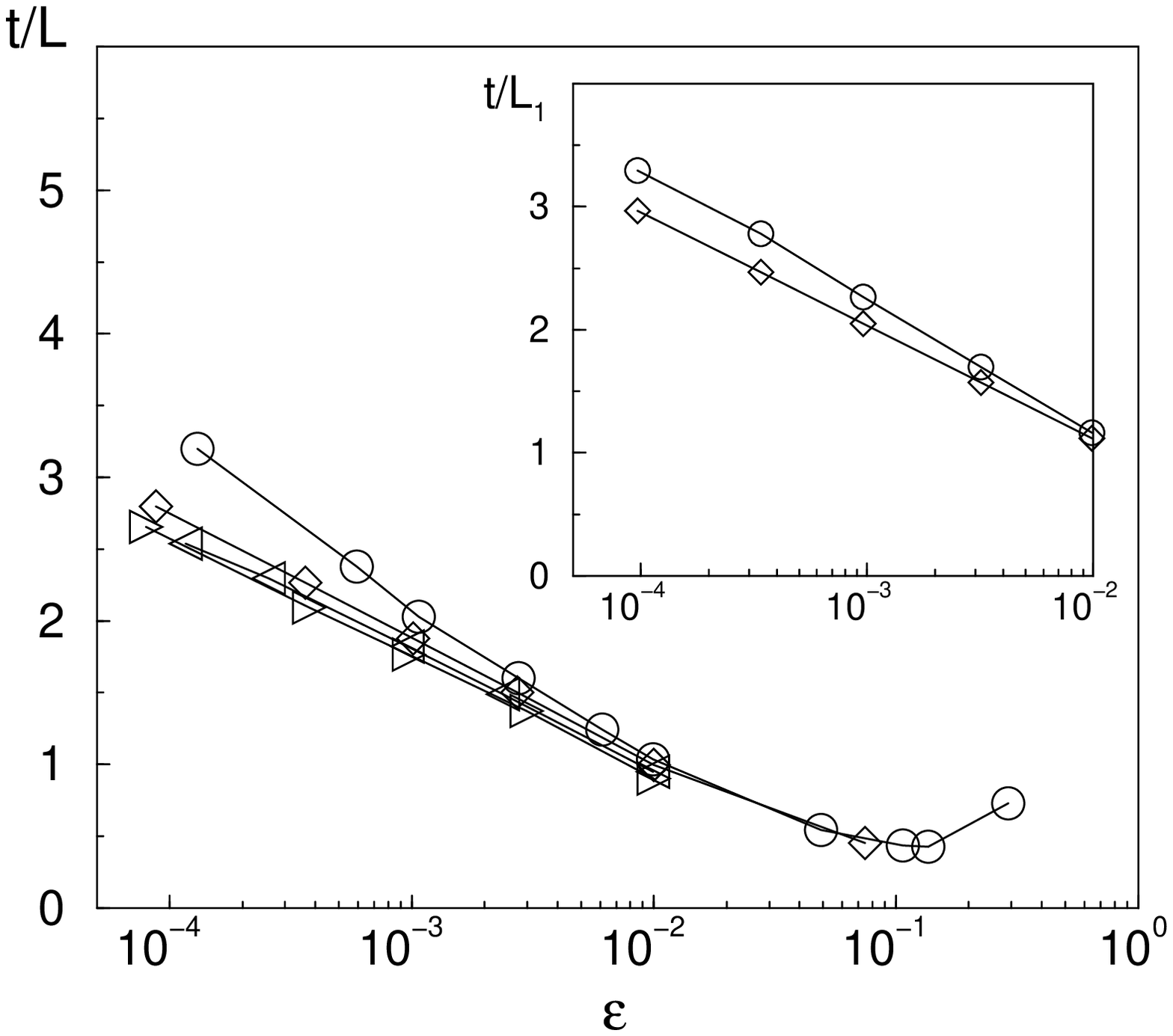}
\caption{\label{teps} K. Shundyak and R. van Roij}
\end{figure}

\begin{figure}[h]\centering
\includegraphics[width=8.5cm]{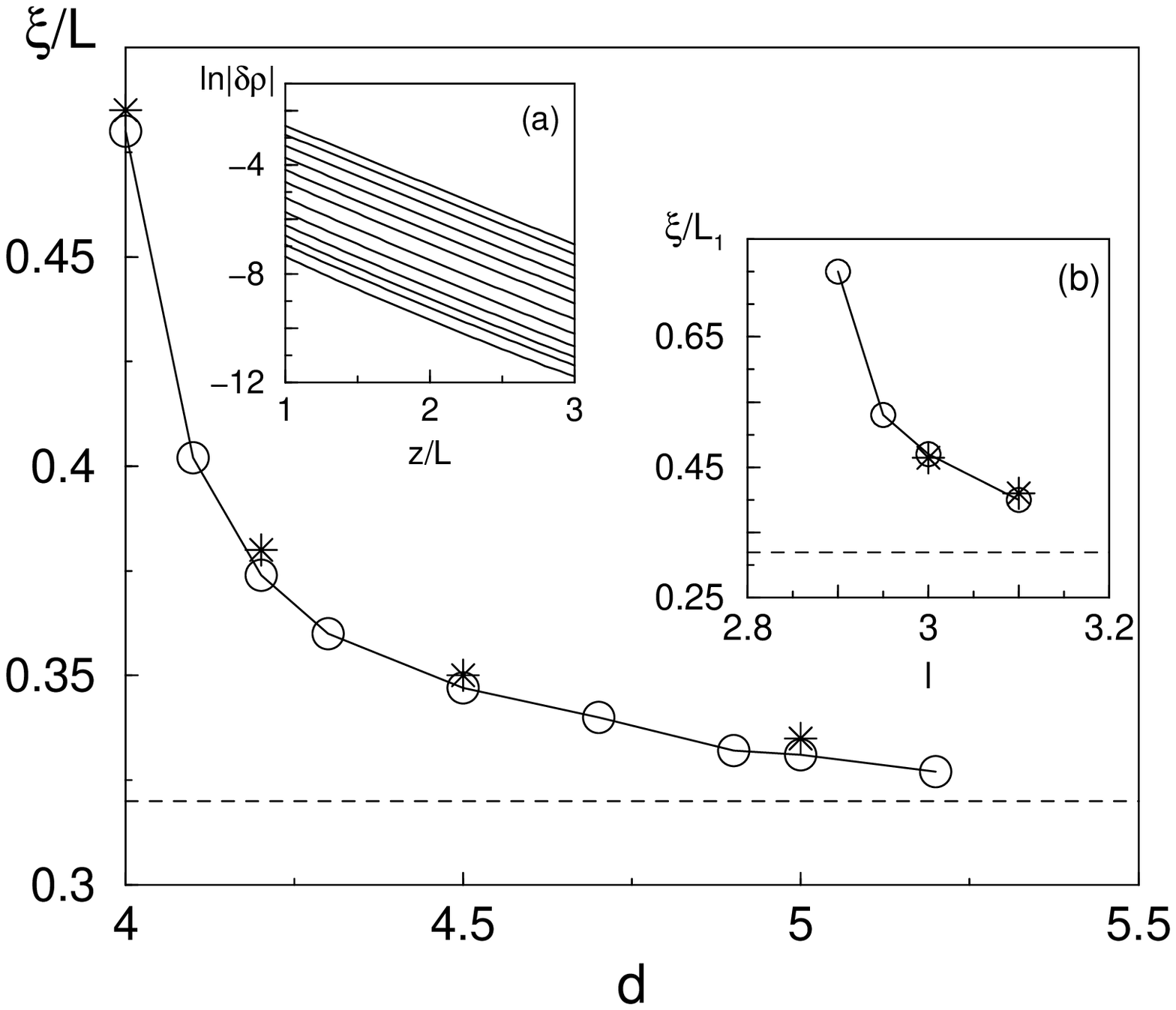}
\caption{\label{xid} K. Shundyak and R. van Roij}
\end{figure}

For all explored mixtures the thickness of the interface $t/L$ (or
the adsorption) was found to diverge logarithmically with
$\epsilon \rightarrow 0$ as shown in Fig. \ref{teps}. For
short-ranged interactions one expects, on the basis of mean-field
theory \cite{Sh}, that
\begin{eqnarray}
\label{logdivads} t=-\xi \ln(\epsilon)+C,
\end{eqnarray}
where $C$ is an irrelevant constant offset, and $\xi$ is the
correlation length of the wetting film. This implies that the bulk
correlation length $\xi_{N_1}$ of the wetting $N_1$ phase should
follow from the slope of the logarithmic growth of $t/L$ in Fig.
\ref{teps}. The asterisks ($\ast$) in Fig. \ref{xid} show the
resulting values of $\xi_{N_1}$ as a function of $d$. These can be
compared to the correlation length that one can extract from the
asymptotic decay of the one-particle distributions
$\rho_{\sigma}(z,\hat{\omega})$ of the $IN_1$ interface into the
bulk $N_1$ phase. This decay can best be analyzed in terms of the
deviation from the $N_1$ bulk density $\delta \rho_{\sigma}
(z,\hat{\omega}) = \rho_{\sigma} (z,\hat{\omega}) -
\rho_{\sigma}^{N_1}(\hat{\omega})$, which we find to decay as
\begin{eqnarray}
\label{asymptdens} \delta \rho_{\sigma}(z,\hat{\omega}) =
A_{\sigma} (\hat{\omega}) \exp (-z/ \xi_{N_1}),\;\;\;\;
z\rightarrow \infty,
\end{eqnarray}
where the only species and orientation dependence is in the decay
amplitude $A_{\sigma} (\hat{\omega})$, i.e., the decay
(correlation) length $\xi$ is one and the same for all species and
orientations. Such a "decay-law", with a single correlation
length, is well-known in mixtures of simple liquids
\cite{ECHHJCP94}. The form (\ref{asymptdens}) is illustrated for
$d=4.0$ at $p=p_t$ in the inset (a) of Fig. \ref{xid}, where all
curves (representing different $\theta$'s) are parallel on a
logarithmic scale. The correlation length follows from the
(common) slope of these curves, and is marked by ($\circ$) in the
main figure. The agreement with the values of $\xi_{IN_1}$
obtained from the logarithmic growth of the interface thickness is
clearly good. The inset (b) of Fig. \ref{xid} shows the similar
dependence of the bulk correlation length of triple-point $N_1$
phase in the case of long-short mixtures, i.e., as a function of
rods length ratio $l$, using the same symbols.

\begin{figure}[h]\centering
\includegraphics[width=8.5cm]{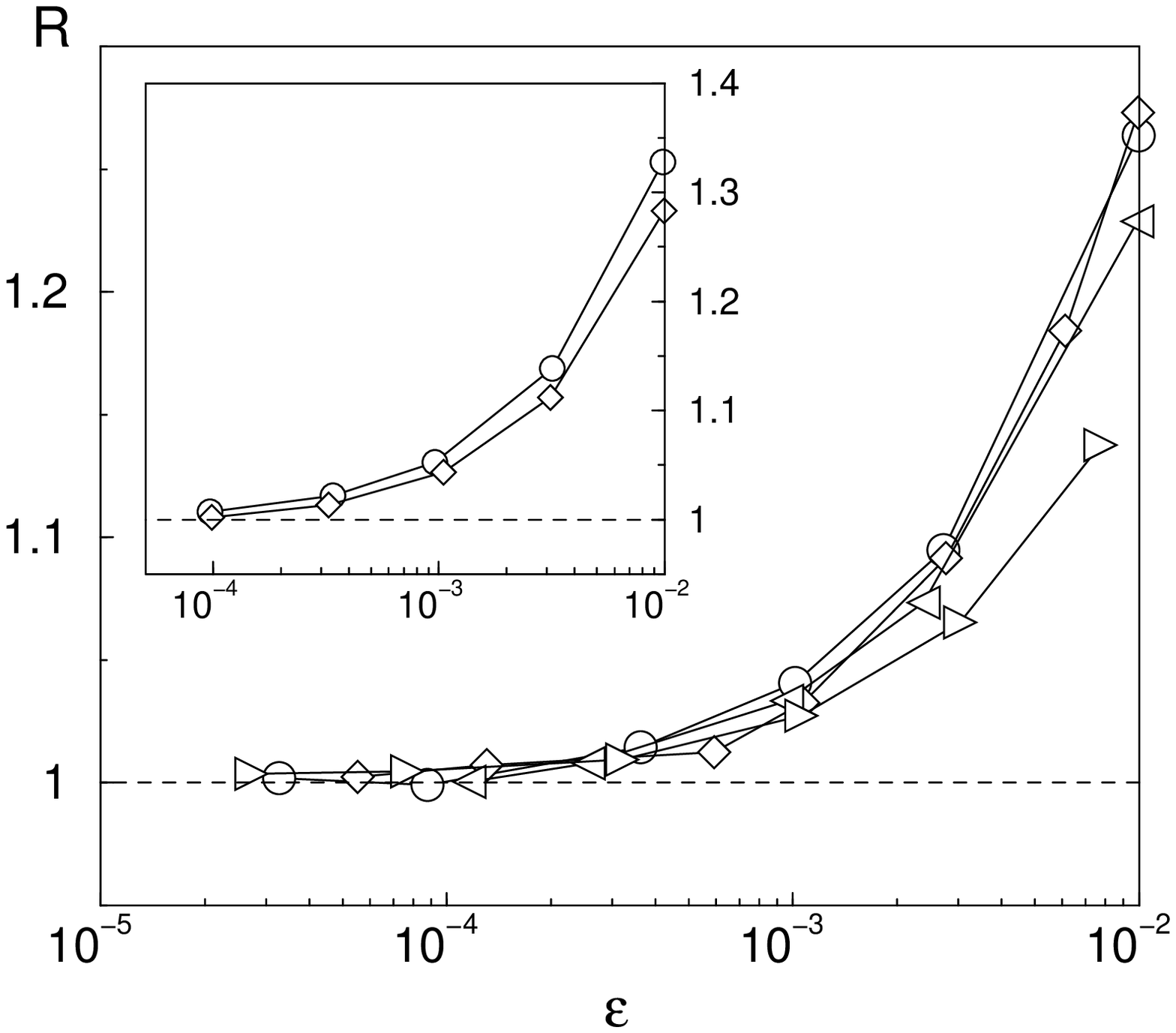}
\caption{\label{Rgraph} K. Shundyak and R. van Roij}
\end{figure}

In order to verify the thermodynamic condition of complete
wetting, $\gamma_{IN_2}=\gamma_{IN_1}+\gamma_{N_1N_2}$ at the
triple-point pressure $p=p_t$, we determine the ratio of surface
tensions
\begin{eqnarray}
\label{R} R(\epsilon)=\frac{\gamma_{IN_2}(\epsilon
)}{\lim_{p\downarrow p_t}(\gamma_{IN_1}+\gamma_{N_1N_2})},
\end{eqnarray}
shown in Fig. \ref{Rgraph}. For all diameter ratios $d$ considered
here $\lim_{\epsilon \rightarrow 0} R(\epsilon)=1$, which implies
a vanishing contact angle. This constitutes the thermodynamic
proof of complete triple point wetting in all thin-thick hard-rod
mixtures with $d \geq 4$. For mixtures of long-short rods the
behavior of $R(\epsilon)$ is the same, as shown in the inset of
Fig. \ref{Rgraph}. A mean-field analysis of the asymptotic
behavior of the surface tension ($\epsilon \rightarrow 0$) in the
case of complete wetting shows that \cite{Sh}
\begin{eqnarray}
R(\epsilon)-1 \sim \epsilon^{2-\alpha}
\end{eqnarray}
with the critical exponent $\alpha=1$. Analysis of our results in
Fig. \ref{Rgraph} gives $\alpha=1.00 \pm 0.05$ which can be
considered as a rough consistency test of our mean-field
calculations.

\section{Summary and discussion}
\label{summarysect} We have studied free interfaces of binary
mixtures of hard rods of either different diameters or different
lengths within Onsager's second virial functional. On the basis of
a vanishingly small wetting parameter, which implies that
capillary fluctuations are not important, we argued that this
mean-field functional provides a realistic description of
isotropic-nematic interfaces of long hard rods. We focussed on
diameter ratios $d>3.8$ (and length ratios $l>3.0$), for which the
bulk phase diagram exhibits an $IN_1N_2$ triple point, and
restricted attention to the case where $\hat{n} \bot \hat{z}$,
with $\hat{n}$ the bulk nematic director and $\hat{z}$ the
interface normal. This is the thermodynamically most favorable
geometry.

We have determined the behavior of the surface tensions of $IN_1$,
$N_1N_2$ and $IN_2$ interfaces between coexisting isotropic and
nematic phases as a function of the bulk pressure and/or the
diameter ratio $d$ (length ratio $l$, respectively). The tension
$\gamma_{IN_1}$ is always very close to the tension of the pure
fluid of thin (or short) rods, and $\gamma_{N_1N_2}$ varies from
zero at the consolute $N_1N_2$ point to values as large as
$\gamma_{IN_1}$ at the triple point for $d\simeq 4-5$. The
thickness of the $IN_1$ and $N_1N_2$ interfaces are always of the
order of the lengths of the rods, except close to the $N_1N_2$
consolute point, of course. The surface tension $\gamma_{IN_2}$ is
found to change non-monotonically with pressure, exhibiting a
maximum close to (but not at) that pressure where the bulk
fractionation is maximal. This maximum surface tension is
considerably larger than that of the pure systems of the
components, typically by a factor of order 3-5, not unlike the
findings of Ref. \cite{SRJPCM2001}, where the case $d=3$ (without
any triple point) was studied. The biaxiality was found to be very
small in all cases, similar to the findings in
Refs.\cite{CPRE93,SRJPCM2001} for the one-component case. Perhaps
our most interesting finding is the phenomenon of complete
triple-point wetting of the $IN_2$ interface by an $N_1$ film. The
thickness of this film is found to diverge as $-\xi\ln(1-p/p_t)$
when $p\rightarrow p_t$, with $\xi$ the correlation length of the
bulk $N_1$ phase, $p_t$ the triple-point pressure, and $p<p_t$ the
pressure. The triple-point wetting phenomenon is confirmed by the
numerical value of the surface tensions, which satisfy $\lim_{p
\uparrow p_t} \gamma_{IN_2}(p)= \gamma_{IN_1}(p_t)+
\gamma_{N_1N_2}(p_t)$. Such a complete wetting scenario was found
for all diameter ratios $3.9<d<5.2$ and length ratios $2.9<l<3.1$
studied here. We expect that this finding will also hold for more
extreme ratios $d>5.2$ and $l>3.1$, which are more difficult to
analyze numerically because of the pronounced nematic ordering of
the triple-point $N_2$ phase.

The predicted phenomenon of triple-point wetting may well be
observable in the experimental system of bare and PEG-coated $fd$
virus particles \cite{PFprivate} mentioned in the Introduction. We
hope that this work stimulates further experimental activities in
this direction.

Another interesting direction for future theoretical work would be
to consider isotropic-nematic interfaces of polydisperse mixtures,
e.g. extending the theory for bulk systems developed in Ref.
\cite{SJPCM02}. For suspensions of length-polydisperse cellulose
experimental measurements of the surface tension have been
performed \cite{CGL02}, and show that the surface tension is much
larger than that of a pure system of rods. It is tentative to
speculate that the fractionation effect that is also present in
binary systems may explain this increase of the tension. We hope
to address this question in future work.

\begin{acknowledgments}
It is a pleasure to thank Marjolein Dijkstra, Henk Lekkerkerker,
and Seth Fraden for stimulating discussions, and Seth Fraden and
Kirstin Purdy for sharing unpublished experimental results with
us. This work is part of the research program of the 'Stichting
voor Fundamenteel Onderzoek der Materie (FOM)', which is
financially supported by the 'Nederlandse organisatie voor
Wetenschappelijk Onderzoek (NWO)'.
\end{acknowledgments}

\appendix

\section*{Appendix}
\label{appendix} In terms of $A=\frac{1}{2} \max(L_{\sigma}
\hat{\omega} \cdot \hat{z}, L_{\sigma'} \hat{\omega}' \cdot
\hat{z})$, $B=\frac{1}{2} \min(L_{\sigma} \hat{\omega} \cdot
\hat{z}, L_{\sigma'} \hat{\omega}' \cdot \hat{z})$ and the
excluded volume $E_{\sigma{\sigma}'}=L_{\sigma}
L_{\sigma'}(D_{\sigma} +D_{\sigma'}) |\sin(\arccos (\hat{\omega}
\cdot \hat{\omega'}))| $, the results of Ref. \cite{PHPRA88}
reduce for $D_{\sigma}/L_{\sigma} \rightarrow 0$ to the following
expression for the "excluded slab" used in Eq. (\ref{rhoreduced})

\begin{widetext}
\begin{eqnarray*}
\mathcal{K}_{\sigma{\sigma}'}(|z|,\hat{\omega},\hat{\omega}')=
\left\{
\begin{array}{ll}
0,\ & |z|>|A|+|B|,\\
\displaystyle\frac{E_{\sigma{\sigma}'}
(\hat{\omega},\hat{\omega}')}{4|AB|}(|A|+|B|-|z|),\ &
|A|-|B|\leq |z|\leq |A|+|B|,\\
\displaystyle\frac{E(\hat{\omega},\hat{\omega'})}{2|A|},& |z|\leq
|(|A|-|B|)|.
\end{array}
\right. \label{Kpon}
\end{eqnarray*}
\end{widetext}

\begin{figure}[h]\centering
\includegraphics[width=8.5cm]{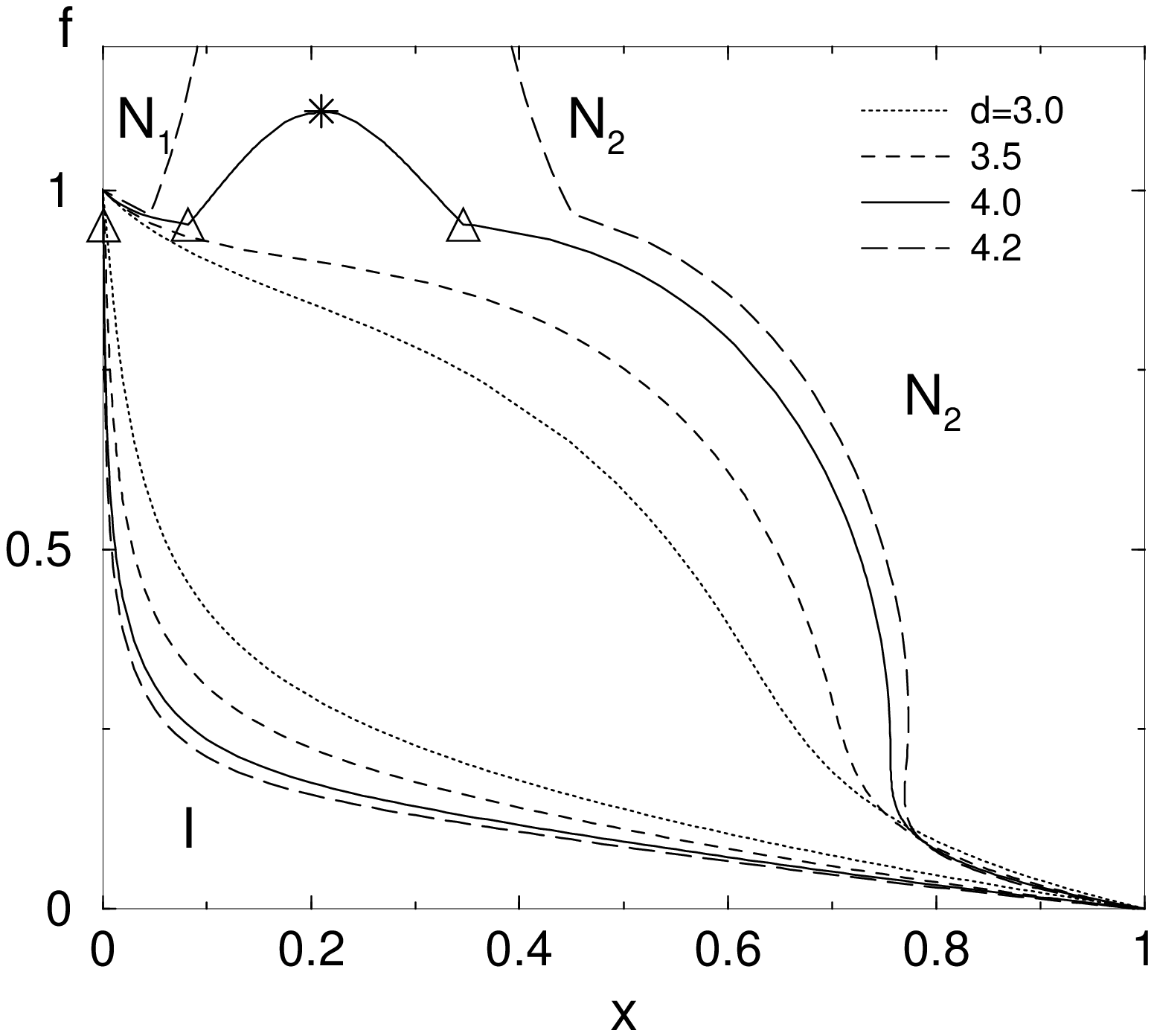}
\includegraphics[width=8.5cm]{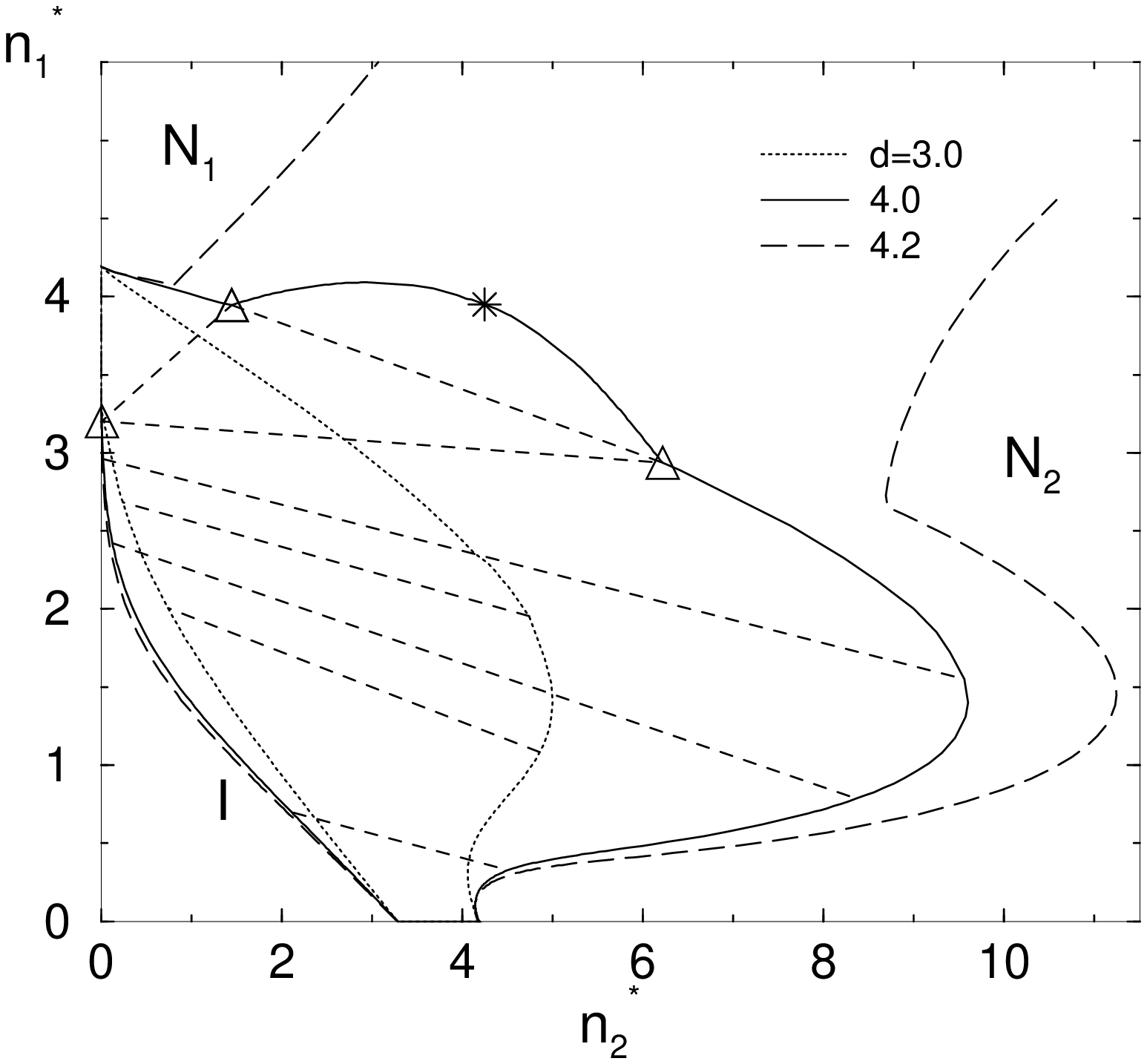}
\caption{\label{thinthickphase} K. Shundyak and R. van Roij}
\end{figure}

\newpage
FIGURE CAPTIONS
\begin{enumerate}
\item
(a) Bulk phase diagrams of binary thin-thick mixtures for
different diameter ratios $d$ in the $f-x$ representation, with
$f=(p-p_{thick})/(p_{thin}-p_{thick})$ the dimensionless shifted
pressure, and $x$ the mole fraction of the thicker rods. We
distinguish the fully symmetric isotropic phase ($I$) and
orientationally ordered nematic phases ($N_1$ and $N_2$). For the
diameter ratio $d=4.0$ the $IN_1N_2$ triple phase coexistence is
marked by ($\triangle$), and the $N_1N_2$ critical point by
($\ast$). (b) The same phase diagrams in density-density
representation, where $n_1^{*}=n_1LD_1^2(\pi/4)$ and
$n_2^{*}=n_2LD_2^2(\pi/4)$ are the dimensionless bulk number
densities of thin and thick rods, respectively. The tie-lines
connect coexisting state points.

\item
Bulk phase diagrams of binary long-short mixtures for different
length ratio $l$ in the $f-x$ representation (see caption to Fig.
\ref{thinthickphase}). ($\triangle$) mark $IN_1N_2$ triple phase
coexistence and ($\ast$) marks $N_1N_2$ critical point for
$l=3.0$.

\item
Dimensionless surface tension $\gamma^{*}=\beta \gamma /LD_1$ of
$IN_1$ ($\circ$) and $N_1N_2$ ($\diamond$) interfaces at triple
phase coexistence ($p=p_t$) for different diameter ratio $d$ of
thin-thick mixtures. The dashed line corresponds to the surface
tension of the one-component $IN$ interface for which
$\gamma_{IN}^{*}=0.156 \pm 0.001$ \cite{SRJPCM2001}.

\item
Biaxiality profiles $\Delta_{\sigma}(z)$ of the thin (without
symbols) and thick (marked by $\times$) rods in the $IN_1$ and
$IN_2$ (at undersaturation $\epsilon=5\cdot 10^{-4}$) interfaces
for diameter ratio $d=4.0$. The inset shows the same quantity for
the $N_1N_2$ interface.

\item
Dimensionless surface tension $\gamma_{IN_2}^{*}=\beta
\gamma_{IN_2} /LD_1$ at $IN_2$ interfaces as a function of
dimensionless pressure $p^{*}=\beta p L^2D_1(\pi/4)$ for different
diameter ratio $d=4.0 (\circ)$, $4.2 (\diamond)$, $4.5
(\triangleleft)$, $5.0 (\triangleright)$ of thin-thick mixtures.
The dashed line indicates the pressure of maximum fractionation.
The data for $d=3.0 (\square)$ are included for comparison (from
Ref. \cite{SRJPCM2001}).

\item
Density profiles of the thin rods $n_1^*(z)$ (a) and the thick
rods $n_2^*(z)$ (b) in the $IN_2$ interface for diameter ratio
$d=4.0$ at triple point undersaturations $\epsilon=1-p/p_t=0.29,
0.1, 0.01, 5 \times 10^{-4}, 1.3\times 10^{-4}, 2.5\times
10^{-5}$. The bulk $I/N_2$ phase is at $z \rightarrow
-\infty/\infty$. The dashed lines $n_1^* =3.977$ and $n_2^*
=0.312$ represent the bulk density of thin (thick) rods in the
triple point $N_1$ phase. These profiles indicate the formation of
a wetting $N_1$ film in the $IN_2$ interface.

\item
Thickness $t/L$ as a function of the undersaturation
$\epsilon=1-p/p_t$ from the triple point pressure $p_t$ for
diameter ratios $d=4.0 (\circ), 4.2 (\diamond), 4.5
(\triangleleft), 5.0 (\triangleright)$  of thin-thick mixtures.
The inset shows the film thickness $t/L_1$ for long-short mixtures
with length ratio $l=3.0 (\circ), 3.1 (\diamond)$.

\item
Correlation length for rods in the triple-point $N_1$ phase of
thin-thick mixtures as a function of diameter ratio $d$,
determined from adsorption ($\ast$) analysis (Eq. \ref{logdivads})
and from density asymptotics ($\circ$) (Eq. \ref{asymptdens}).
Inset (a) shows $\ln|\delta \rho (z,\theta)|$ for several values
of $\theta$ at the $N_1$ side of the $IN_1$ interface for $d=4.0$.
Inset (b) displays the correlation length (in units of length of
short rods $L_1$) for long-short mixtures as a function of length
ratio $l$. The dashed lines indicate the correlation length for
rods in the $N$ phase of the monodisperse hard-rod fluid.

\item
Surface tension ratio $R$ [see Eq. (\ref{R})] as a function of the
triple point under-saturation $\epsilon$ for diameter ratio $d=4.0
(\circ), 4.2 (\diamond), 4.5 (\triangleleft), 5.0
(\triangleright)$. The inset shows the same quantity for
long-short mixtures $l=3.0(\circ ), 3.1 (\diamond )$.

\end{enumerate}

\newpage

\newpage

\newpage

\newpage

\newpage

\newpage

\newpage

\newpage

\newpage

\end{document}